\documentclass{article}
\usepackage[utf8]{inputenc}
\usepackage[english]{babel}
\usepackage{todonotes}
\usepackage{ulem}
\usepackage{csquotes}
\usepackage{url}

\usepackage{natbib}
\setcitestyle{authoryear,open={(},close={)}}

\usepackage{geometry}
\title{\textbf{
Artificial Intelligence and Aesthetic Judgment} \\

}

\author{Jessica Hullman\footnote{Department of Computer Science, Northwestern University.} \and Ari Holtzman\footnote{Department of Computer Science, University of Chicago.} \and Andrew Gelman\footnote{Department of Statistics and Department of Political Science, Columbia University.}  
}
\date{19 Aug 2023}

\begin{document}

\maketitle
\begin{abstract}
Generative AIs produce creative outputs in the style of human expression. 
We argue that encounters with the outputs of modern generative AI models are mediated by the same kinds of aesthetic judgments that organize our interactions with artwork. 
The interpretation procedure we use on art we find in museums is not an innate human faculty, but one developed over history by disciplines such as art history and art criticism to fulfill certain social functions. This gives us pause when considering our reactions to generative AI, how we should approach this new medium, and why generative AI seems to incite so much fear about the future.
We naturally inherit a conundrum of causal inference from the history of art: a work can be read as a symptom of the cultural conditions that influenced its creation while simultaneously being framed as a timeless, seemingly acausal distillation of an eternal human condition. In this essay, we focus on an unresolved tension when we bring this dilemma to bear in the context of generative AI: are we looking for proof that generated media reflects something about the conditions that created it or some eternal human essence? Are current modes of interpretation sufficient for this task? Historically, new forms of art have changed how art is interpreted, with such influence used as evidence that a work of art has touched some essential human truth. As generative AI influences contemporary aesthetic judgment we outline some of the pitfalls and traps in attempting to scrutinize what AI generated media \textit{means}.

\end{abstract}

\section{Introduction}

\begin{displayquote}
``If a man who was capable by his cunning of assuming every kind of shape and imitating all things should arrive in our city, bringing with him the poems which he wished to exhibit, we should fall down and worship him as a holy and wondrous and delightful creature, but should say to him that there is no man of that kind among us in our city, nor is it lawful for such a man to arise among us, and we should send him away to another city, after pouring myrrh down over his head and crowning him with fillets of wood.'' --- Plato, {\em The Republic}, translated by \citet{shorey1937republic}. 
\end{displayquote}

Chatbots have made interacting with artificial intelligence accessible to a much broader swath of society. There is a feeling among many experts as well as the general public that we are witnessing a breakthrough, following rapid progress in the performance of deep generative models due to new algorithms and increases in scale of models and training sets. In contrast to technological achievements such as nuclear weapons that came after scientists had developed new physical theories, recent advances in deep learning have arisen in the absence of formal frameworks that can explain exactly how such success is achieved. A flurry of recent work in machine learning and related fields has scrambled to understand how large generative models such as GPT-4 learn and reason and in what ways these processes resemble human cognition \citep{bubeck2023sparks,olsson2022context,wei2022chain,wei2022emergent}.
There have been calls for a new science of intelligence~\citep{mitchell2023debate} and interdisciplinary studies of machine behavior~\citep{rahwan2022machine}. 

At the same time, the perceived humanness of deep model outputs has incited a wave of criticism of what is suggested to be an overhyped landscape of AI research, and scholarly communities such as FAccT\footnote{ACM Conference on Fairness, Accountability, and Transparency (ACM FAccT), \url{https://facctconference.org/index.html}} and AIES\footnote{AAAI/ACM Conference on Artificial Intelligence, Ethics, and Society (AIES), \url{https://www.aies-conference.com/2023/}.} have organized around discussing the harms of large generative AIs.  
The dangers of attributing consciousness or general intelligence to these systems is at the forefront of these discussions, with critics suggesting that the concept of artificial general intelligence is vague at best, and unobtainable~\citep{bishop2021artificial,vanrooij2023} and associated with social movements like eugenics at worst~\citep{gebru_eugenics}.  
Conversely, some within the research community have argued that a fear of hype on the part of experts has led to recent alarmism around generative AI, as the research literature had tended to underclaim what language models were capable of~\citep{bowman2022dangers}, limiting our ability to prepare for the potentially enormous impacts of future advances.

Concerns with the aims and significance of AI may seem to have little to do with human aesthetic production. How to make sense of a painting or sculpture hardly seems as urgent as figuring out what to make of recent advances in AI. Comparing the influence of art on society to the potential influence of generative AIs can seem downright inappropriate. After all, while a piece of art might be delightful just like a whimsical poem or artistically-rendered image from a generative AI can be delightful, it's hard to imagine how experiencing art in a museum, or the soundtrack to a movie or a popular piece of fiction for that matter, could be as damaging as a decision-maker taking at face value a factually incorrect or deceptive response from a chatbot. 
Instead, much of the recent discourse on generative AI and art explores disruption to human artists, including methodologies~\citep{shan2023glaze, gandikota2023erasing} and legal questions of ownership and credit and the future of creative work and the media enterprise~\citep{epstein2023art}. 
There is a sense that AI-generated creative work, like so many other instances of AI in modern life, threatens to strip power from human producers, and in light of these 
changes, we must preserve the long-running institutions of art. 

However, it is easy to overlook how our understanding of how works of art carry meaning, and what it means to stand before a piece and put it in context (with the help of museum placards or art critics), have always reflected the needs of the historical moment. 
Modern conceptions of how to judge a work of art are the result of a series of developments in cultural history, induced by fears that coincide with changes in our ability to represent the world much like those we see reflected today in media around generative AI.
To understand how we got from Plato's bizarre fear of the poet to a modern notion of taste in art driven by the perceived absence of artificiality rather than the presence of some particular quality, we should acknowledge that the ways we attribute meaning to aesthetic objects have been shaped, and will continue to be shaped, by cultural and political agendas as well as existential crises.  

Today's attempts to judge whether AI achieves something inherit aesthetic standards handed down by the history of art. 
In judgments of both art and generative AI we see an urgency around how to make sense of developments in aesthetic production from a teleological viewpoint that we hope will help us anticipate future developments in this fast-moving technology. There is a tendency to locate the essence of humanness via reference to what a work does not contain, and to define what is human as the opposite of the algorithmic.  
Acknowledging these differences casts the urgently-issued philosophical judgments of generative AI we see today in a different light, as seemingly parochial and archaic symptoms of the blurry boundaries we draw around acts of aesthetic production. 
As technology capable of producing aesthetic objects changes, we should expect the future of our interactions with AI to be intertwined with the evolution of aesthetic judgment.

Understanding judgments of AI through the lens of judgments about art also exposes weaknesses in the new philosophical arguments we are working to build. 
On a semiotic level, our judgments of AI realize the same \textit{displacement of causality} that results from combining both metonymic and metaphoric modes of reading objects in a single system of interpretation. 
We find it hard not to assume that changes in the form of a creative product (or a lack thereof) correspond to changes in beliefs, attitudes, intentions or social, political, cultural conditions (or a lack thereof).
Objects are read as ``signals'' of underlying states of the world, whether at the level of technical details behind their construction (e.g., training technique X causes distribution Y in outputs) or social factors behind their creation (e.g., lack of ethical commitments A, B, C cause biased application Z of outputs). 
At the same time, we acknowledge that we don't know what we are looking for, because the target of our work (``human-like intelligence'') is itself believed to be transcendent.
And so our attempts at causal attribution can at any time be eclipsed by reference to some emergent universal human principle located outside of any particular object. In other words, if we simply assume—without examination—that whatever we are seeking in media is lacking in AI generated media, analyzing what AI generated media \textit{is} doing becomes self-defeating.

\section{What do we expect when we judge generative AIs?}
In a recent magazine article, 
Douglas Hofstadter, computer scientist and author of \textit{Gödel,~Escher,~Bach}, considers a mock first-person text that ChatGPT was prompted to generate to explain his reasons for writing the acclaimed work~\citep{hofstadter1999godel,hofstadter2023atlantic}.
In several seemingly eloquent paragraphs, 
ChatGPT's response describes how the author, long fascinated by ``the deep connections that underlie seemingly unrelated fields,'' found himself enthralled with themes of self-reference and recursion in the works of the three men, then realized that the unifying thread was not just these themes, but how ``[e]ach of them pushed the boundaries of what was thought possible, revealing deep truths about the nature of reality, the limits of human knowledge, and the beauty that can emerge from complexity.''

Hofstadter derides the response, dissecting several sentences like the above that, despite appearing at first to be ``convincing and true'' or ``noble sounding,'' are instead ``vague,''``vapid pablum'' that could not have been authored by him. 
He recounts his actual reasons for writing the book, pointing out how GPT-4's responses fails to reference key aspects of the original text such the use of vivid analogies, humor, and musically-influenced dialogues.
He calls it baffling that anyone would let opaque computational systems perform intellectual tasks, and that ``when it comes to using language in a sensitive manner and talking about real-life situations'' it makes no sense whatsoever to be using chatbots; in fact the ``artificiality of the creation runs counter to [his] lifelong belief system.''
He warns repeatedly of the dangerous illusion that even ``highly intelligent'' and ``unquestionably insightful'' people may fall victim to, concluding that if falling for this illusion happens sufficiently often and becomes accepted, it will ``undermine the very nature of truth on which our society---and I mean all of human society---is based.''

What do we learn from Hofstadter's essay?
While his comments on how GPT-4 got the story of his book wrong are unique, the critical stance of his essay resemble perspectives on generative AI we are seeing frequently from intellectual figures and ordinary citizens alike. Specific evidence varies (i.e., the prompt given to the model), but
we seem to be witnessing the same beauty contest replaying everywhere we look, where the outputs of a generative model like GPT-4 are held against a human standard and deemed to fall far short of the mark.
We are warned again and again that they cannot be authentic or grounded the way human speech is, and that the technology is moving us toward a ghastly future in which we are deluded by representations that we ourselves brought into being.

The goal of this essay is not to debate these points.  Instead, we ask: How unique are these sentiments as judgments of human-like creative production? 
We can start by considering what sort of ``truth'' we might be seeking in the output of a generative AI, and what underlying interpretive framework supports the confidence with which many of us dismiss ChatGPT's attempts at reflecting human experience comes. 

\subsection{Metonymic and metaphoric meaning}
Reflect for a moment on the frame of mind most of us so naturally assume when we stand before a piece of art in a museum setting and try to interpret its significance. Many of us dutifully read the placards next to each piece, which help us contextualize the work by locating the work and its author within within a broader artistic geneology. At the same time, we are encouraged to see beyond the specific materials and their historical relevance to the less tangible ``essence'' of the work, to sense the elusive truth that renders museum-worthy those works that achieve a certain authenticity of expression. This kind of dual orientation, where meaning is read into works of art via both metonymic and metaphorical relationships, is so ingrained that we hardly stop to question it~\citep{preziosi2003brain}.

Hofstadter does not approve of GPT-4 being used to overwrite his authentic experience in creating the book as author. 
The idea that a narrative other than the true story of the conditions behind a work could be used to gain insight into it is incompatible with a tendency to see works of art (or in this case, artful nonfiction) as unique specimens, or evidence of, the conditions under which they were created.
Each work of art reflects a sort of location in a network of such objects, taken as a meaningful representation of the conditions under which it was created. 
The very act of asking ChatGPT to describe Hofstadter's motivation for his book violates an unwritten assumption that works of art are meant to be read as reports back from some specific circumstances or scene, and cannot be separated from these conditions without a loss of authenticity. We take for granted that the more mundane historical facts on that temporal spatial location are not sufficient to tell us about the sensibilities of the author or the time of the work's creation.
And so, the lack of definitive origin point of ChatGPT's response prevents it from carrying meaning within this interpretive framework.

Consider the diffuse nature of authorship in generative AI, which has caused a feeling of uneasiness in the discussion of generative AI and art. If the prompter is not the author of a work then what does the work reflect? And what is our procedure for agreeing on its intent and meaning? \citet{foucault2017author} points out that the purpose of an author is determined by the interpretive norms of the discourse a work inhabits—and many works have not required authors, such as epic poems handed down by oral recitation. What, then, is the function of identifying the origin of elements and themes produced by generative AI? Authorship is often used to determine what readers should be allowed to take away: ``To give an Author to a text is to \dots furnish it with a final signification, to close the writing'' \citep{barthes1967death}. The search for a notion of authorship in AI-generated media is a power move, an attempt to determine what the definitive interpretation of a work is, what should be made canonical. We would like---or perhaps more accurately, we feel we need---to make AI-generated media legible to current interpretive modes, the way governments insist on predefined categories for businesses or marital relationships in order to make human organization legible to the state \citep{scott2020seeing}.

At a higher level, consider how it has become common to contrast differences between the ``origins'' of models created at different times by different groups---in the form of details about their training data and its scale~\citep{denton2021genealogy}, training process~\citep{bengio2013representation}, or model size in parameters~\citep{bommasani2021opportunities}---to locate them in the broader landscape of the generative AI ``movement''
From this metonymic perspective on meaning, legibility is perceived as necessary for the works to function productively in the broader scientific discourse, motivating a plethora of calls for different types of transparency around the generating processes behind the models \citep[e.g.,][]{gebru2021datasheets,haibe2020transparency,mitchell2019model}.
Similar to how art historians interpret the properties of artistic techniques as evidence of the mentality of a culture, e.g., what the subdued color schemes and sharp finish of Dutch still life paintings (not to mention the objects portrayed say about the society's moral culture and wealth~\citep{berger2011caterpillage}, critical scholars develop theories of what general properties of models pursued in deep learning research, like scale and efficiency, say about the values of the cultures that create them~\citep{birhane2022values}.
These details, as well as others like how the models are culturally ``staged'' for consumption, are not treated as mere historical facts, but as signals that inform of the values of the groups behind their creation, grounded in the enabling assumption of art history: that changes in form correspond directly or indirectly with changes in beliefs, mentalities, or intentions, or social, political, or cultural conditions~\citep{preziosi2003brain} (p.96).

However, it is not only the fact that GPT-4's response derives from a source other than Hofstadter that renders its words to be neither ``noble'' nor ``true'' in his eyes, lacking of ``the authenticity of human speech,'' so much that its ``artificiality \dots runs counter to [Hofstadter's] lifelong belief system.''
To find a belief system capable of judging the authenticity of a human expression of an idea or inspiration, we need look only as far as the attitudes we bring to interpreting art.  
Anyone who has gone to an art museum has probably felt compelled, in one way or another, to reflect on the human condition more broadly. In art we look for a property which museums, by housing and juxtaposing different pieces from different cultural locations under a single roof, frame as universally true and somehow especially accessible through art.
One could argue that a painting or sculpture or great work of literature, much more than a tool or work of science from the same population, seems to capture in an exalted way the enduring essence of some singular dimension of ``humanness.''
So great is this yearning to find the truth, in fact, that over-interpretation is seen as implicitly warranted: no matter if the artist themselves insists that a choice of color, or phrasing, was unintentional, it becomes our duty as spectators to read into, or beyond, the work itself for the insight it provides into the human condition, writ broadly.
Ironically, within this system, 
there is a sense in which even Hofstadter's account of the significance of his book cannot really reveal the significance of the work.

We glimpse this kind of metaphorical reading of generative AI whenever we give in to the temptation to read the output as an attempt at some ideal output achievable by human intelligence or consciousness.
If art history is an attempt at 
a ``universal and systematic human science''~\citep{preziosi2003brain}, so too is the study of generative AI, where the outputs are treated as naturally symbolic of human-like intelligence.
The pursuit of ``artificial general intelligence''~\citep{goertzel2014artificial,grudin2019chatbots} in LLM outputs is like the pursuit of the human spirit in works of art.
Just like it would seem unnatural to take in a piece of art without trying to judge how well it has achieved some target human ``essence,'' 
so too it would seem unnatural to try to separate our natural orientation to treat our interactions with the latest generative AIs, offered for interaction via human-like dialogues, as Turing tests. 
But in contrast to the original intent of the test, today's encounters with generative AIs would often seem to be a kind of \textit{aestheticized} Turing test:  rather than merely asking, ``Could something like this be produced by a human?'' we look in the output of a generative AI for signals of "humanness" as defined some internal idealization we presume ourselves to have access to.  
We can't help but expect the exaltation of human intelligence from AI like we expect the exaltation of the human spirit from art. When what we see falls short, as it does so painfully to Hofstadter, we feel the same sort of aesthetic dismay that the art critic feels when faced with something tacky.

The fact that aesthetic judgment depends on several seemingly contradictory modes of meaning-making makes it much of an art than a science. Yet it is an art which often fools us into thinking we are observing reality directly. 
Art itself is a kind of trompe l'oeil, and we cannot easily predict what we are looking for in the illusions it offers. 
Emergence, in the sense of the effect of a whole that seems greater than the sum of its parts, is present in all artistic representation, even if our immersion within the system prevents us from noticing it. 
As linguist Michel \cite{breal1868idees} wrote \citep[as cited by][p.100]{preziosi2003brain}: ``Our eyes think they perceive contrasts of light and shade, on a canvas lit all over by the same light. They see depths, where everything is on the same plane. If we approach a few more steps, the lines we thought we recognized break up and disappear, and in place of differently illuminated objects we find only layers of color congealed on the canvas and trails of brightly colored dots, adjacent to one another but not joined up. But as soon as we step back again, our sight, yielding to long habit, blends the colors, distributes the light, puts the features together again, and recognizes the work of the artist.''

Similarly, perceived breakthroughs in art tend to coincide with the surpassing of what were seen as hard limits on the artist's representational ability. 
The technological sea change brought about by the photo-realism achieved by Impressionist painting, for example, showed that a more ``scientific'' use of light could lead to outputs that were previously impossible to create in two dimensions.
A culture's approach to art is thought to be intimately tied to how they perceive the world; what we consider fine art is much more a technology of vision, achieved through artistic representation, than attempt at mere decoration or aesthetic pleasure through visual elements.

Hofstadter acknowledges, as so many others do, the ``stunning progress'' of ``astoundingly virtuosic'' generative models like GPT-4.  
As in art, witnessing emergence~\citep{wei2022emergent} in the outputs of generative AI systems---here referring to examples where the behavior of a system is implicitly induced rather than explicitly
constructed~\citep{bommasani2021opportunities}---is a key driver of our awe.
For example, ``grokking''~\citep{power2022grokking} refers to how continuing to train a neural net past the point of overfitting to the training data sometimes leads to improvements in validation accuracy toward perfect generalization. Chain-of-thought reasoning refers to how the accuracy of very large models can increase sharply when they are prompted to decompose certain tasks into intermediate steps ~\citep{wei2022emergent}.
In both art and generative AI, the societal reaction to representational breakthroughs is a sense of wonder at an unveiling of reality through the mastery over form and material achieved by the creator.
These are the revelations celebrated by the modern art museum, which curates the story of a triumphant human rendering of reality.
These days, we marvel at modern machine learning as universal function approximator~\citep{hornik1989multilayer}, capable of reconstructing arbitrarily complex phenomena, 
as evidence that we have overcome the constraints on more classical statistical modeling as a means of representing the world.
And yet in art and AI it isn't actually reality that is unveiled, but reality as sign, constructed in a system of expectations about object legibility.

And so, in a system that depends on our becoming convinced that the work brings us closer to reality itself, we 
can point to the techniques used to achieve an effect, but time and time again we overlook the historically contingent, socially constructed nature of the perspective on reality afforded by a new technology of vision. 
Consider a recent ad campaign by Nikon that fights back against the use of generative AIs to produce photorealistic images by featuring stunning real photographs overlaid with prompts that could hypothetically be used to prompt a generative AI~\citep{zhang2023nikon}. 
The slogan reads, Don't give up on the real world.
In machine learning, overlooking our dependence on regimes of representation can manifest, for example, as a forgetfulness when it comes to questioning the representational status of the measurements on which models are trained~\citep{jacobs2021measurement}.

\begin{figure}
    \centering
    \includegraphics[width=0.6\linewidth]{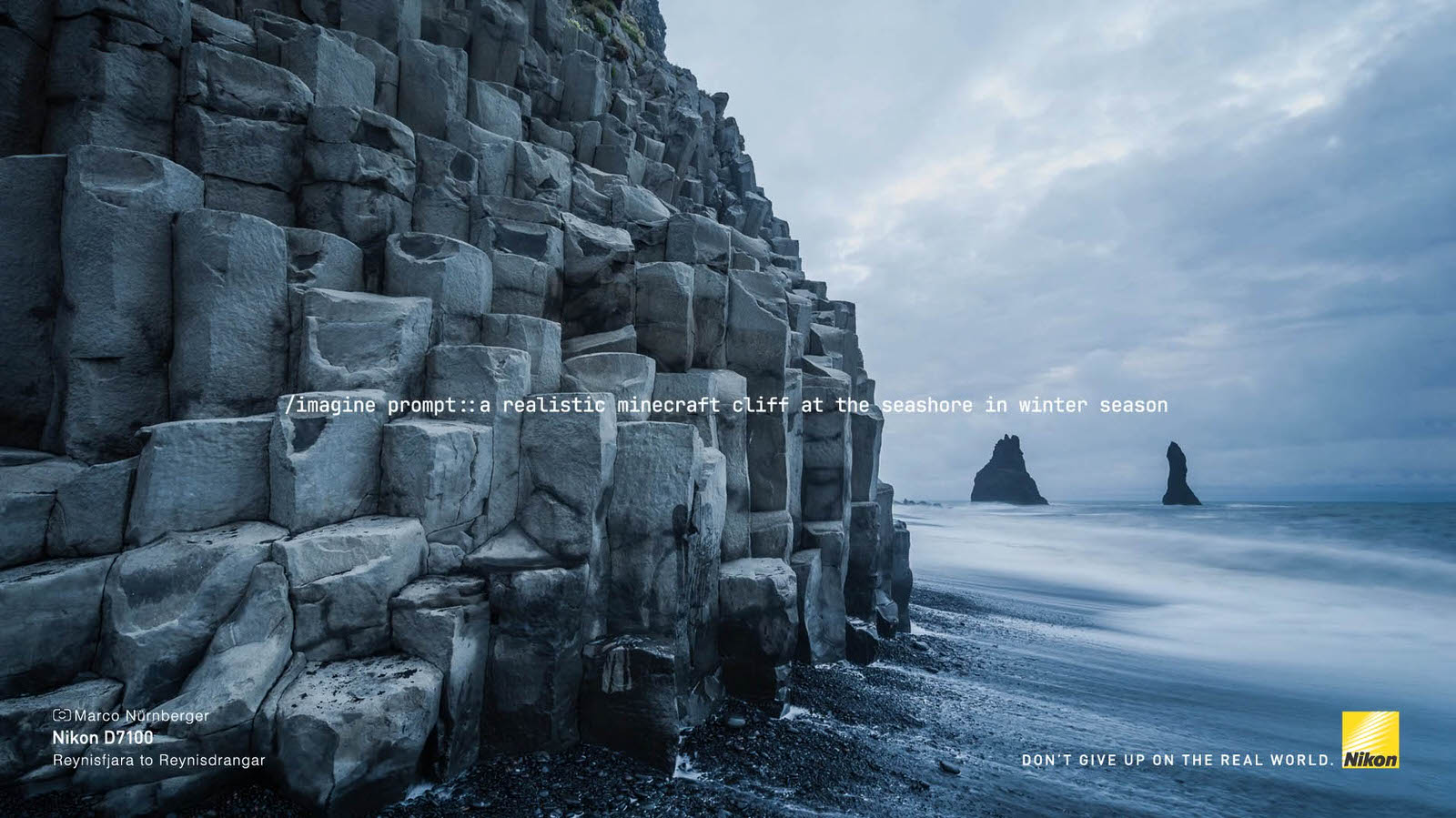}
    \caption{A Nikon ad mocks the rush to generate photorealistic imagery with AI by overlaying real photographs with hypothetical AI prompts~\citep{zhang2023nikon}.}
    \label{fig:nikon}
\end{figure}

\subsection{Decorum}

The staging of art by art history and museology imbues objects with a kind of decorum that makes them seem ``legible in a disciplined manner, construable semiotically as emblems, simulacra, or object lessons; as illustrating or representing desirable or undesirable social relations''~\citep{preziosi2003brain} (p.\ 39).
Hence, we have come to associate legibility with a sense of scrupulous and tasteful self-possession on the part of the artwork.  
In the absence of such decorum, we become particularly conscious of how an object is \textit{not} fulfilling our expectations of legibility. 
Could this be the sense of nobility that Hofstadter cannot find in GPT-4's words?

Perceptions of ChatGPT as a ``king of pastiche''~\citep{klein2023interview} that amounts to glorified cutting and pasting express the disappointment that results from trying to read an object that is ignorant of its own process of becoming, which does not allow the glimpse of the hidden truth of the modern self as we’ve come to expect from works of art.
When an LLM articulates a complex concept as well as we think an expert could have, or displays what strikes us as self-awareness, or responds to our prompt with a witty poem, there's a sense of witnessing the same sort of irreducible genius we expect of a work of art.
And yet, the presence of small discrepancies from our expectations leave us with a vexing sense of unfinishedness in light of the self-possessed fullness or being-at-work that we have come to expect of aesthetic objects.
Sometimes called the "uncanny valley"~\citep{mori2012uncanny}, this phenomenon reflects our sensitivity to deviations from deeply-ingrained aesthetic standards.

Interestingly, it was not until works of art became objects for staging in a museum in the mid-1800s that they became invested with the kind of decorum we expect of great works today.
In many ways, searching for traces of the elusive human essence in the outputs of modern generative AIs is like visiting the Wunderkammer, or ``cabinet of curiosities," that arose in mid-sixteenth century Europe as a precursor to the modern museum, but before the distinction between "fine art" and natural or man-made curiosity  solidified.
The Wunderkammer was very much a realization of pastiche, a visual conversation that arose from the crowded juxtaposition of disparate objects--skulls, taxidermy, paintings, scientific instruments, relics of mythical beasts, preserved plants, and so on. 
With its crammed shelves, the Wunderkammer was a ``conversational jumble of things'' \citep[p.97]{preziosi2003brain}, a ``perplexing organization''~\citep{bennett1998pedagogic} that has been likened to ``disorganized talk.'' 
Today's awkward dialogues with LLMs or deep generative vision models could be said to be a sort of modern-day metaphorical equivalent. 
With a model like ChatGPT we get glimpses of chains of reasoning, or poetic form, or factual expositions that rival curated Wikipedia pages, but along with them sentences that trail off, or loop~\citep{holtzman2019curious}, or made-up references to prior scholarly works.
The insensitivity to meaning when we least expect it, like when an LLM substitutes one famous person or elite university for another, as if it doesn't matter, is like the juxtaposition of rhinoceros feces and the exquisite marble bust of a Roman king.  
In light of our deeply ingrained expectations of the decorum of the aesthetic object, this feels like a violation, just like a restored fresco where the face of the Madonna has come to resemble a monkey~\citep{willis} feels like a violation.
That our interactions with a generative AI can seem to have the same sort of heightened status as our interactions with art works, 
but at the same time demonstrate how far the technology is from being in possession of its own creative purpose, and the aura of decorum and semantic completeness we come to expect in objects staged for aesthetic judgment, makes us uncomfortable.

\begin{figure}
    \centering
    \includegraphics[width=0.5\linewidth]{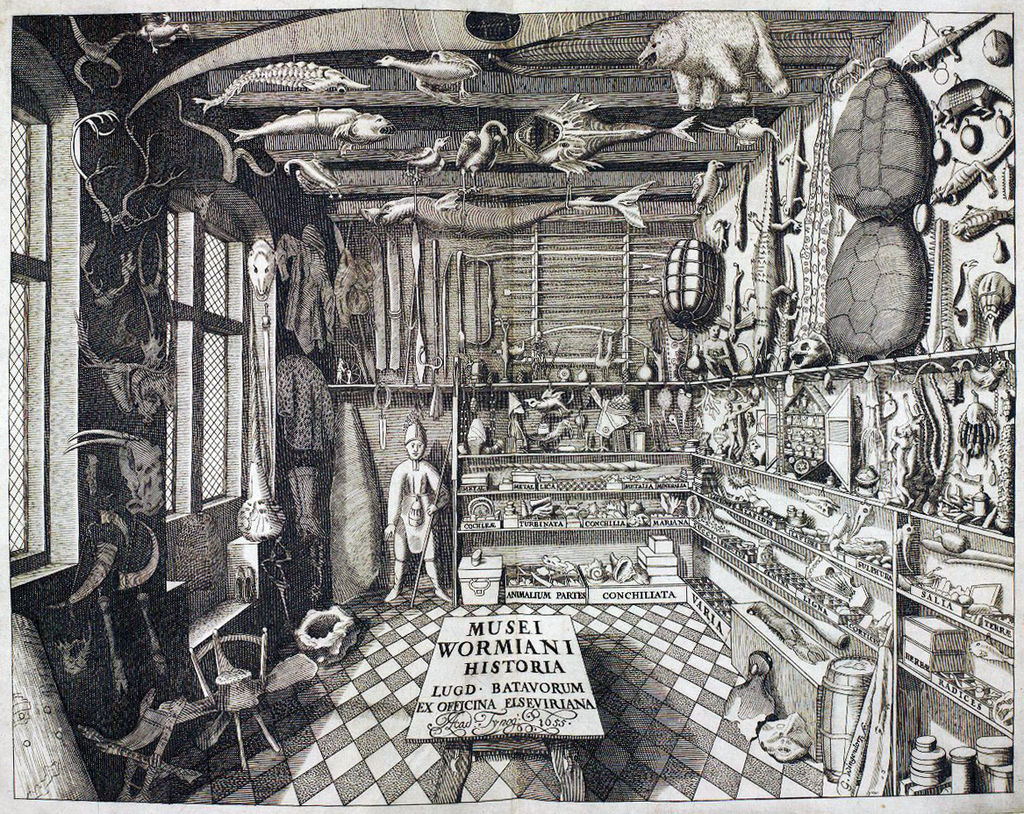}
    \caption{A sketch of a Wunderkammer or Cabinet of Curiosities, precursor to the modern art museum.}
    \label{fig:wunderkammer}
\end{figure}

Consider too the relationship between the output of a generative AI and negative space. Negative space has become a hallmark of the modern museum that arose in Europe in the 1800's as a means of incorporating collections of artworks into ``a seamless narrative and controlling taxonomy,'' relieving the viewer of the work required to make sense of the mess (\citet{stafford1994artful}, as cited in \citet{bennett1998pedagogic}).
The ample white space between works on the walls, and the emptiness created by the high ceilings of the galleries and halls in the modern museum, help reflect the significance of the works inside, while obliquely reminding of those left outside because they fell short of the mark.
Going a step further, the work accomplished by aesthetic judgment comes to orient great art with an absence of something, and in contemporary times we should expect our most exalted interactions with art to leave us with the sense that art's reality is independent of any specific form or content. Duchamp's fountain, for example, achieves its significance as art through a blithe gesture toward that which art is not (product).
Our interactions with generative AIs, on the other hand, put us face to face with input-output machines that are not capable of abstaining, that will fill whatever space we give them, and are therefore barred from using negative space in a way that has come to characterize the cultural aesthetic attached to the notion of ``genius'' \citep{sontag1969aesthetics}. 
That we try to judge these machines by the same sort of standards we bring to works of art reflects a category error, one that continue to distract us as long as we allow our internal aesthetic compass to lead.

\subsection{A teleological perspective}
To understand how our expectations of aesthetic objects are dashed by models like ChatGPT, we must also acknowledge the \textit{narrative} nature of the legibility we expect from great works of human creation. These expectations owe much to the creation of the modern museum as a physical manifestation of the interpretive system we take for granted.
An important aspect of art history's value as ``an organizing and orienting concept,'' one that makes certain notions of the subject ``visible and therefore legible and intelligible'' is that it provides a carefully curated perspective~\citep[p.67]{preziosi2003brain}, which we are always attempting to center ourselves in when we interact with individual works.

More specifically, the perspective that the industries of art history and museology afford depends on a \textit{teleological} orientation, in which the creative principle behind art is read as an inevitable evolution over time toward some ideal state.
The goal of the spectator is to find the point of resolution in which this``natural'' progression, or ``rationalized genealogy'' can be seen clearly, and all things that previously seemed different are resolved into some single truth~\citep{preziosi2003brain}. For example, historians have pointed to how despite their differences, the works of Italian Futurists, German Fauvists, and Spanish Cubists all represent steps toward dissolving the rules of traditional formalist European painting that preceded it.

Now consider the staging of a generative AI model like ChatGPT via online demos. 
Hofstadter's use of his encounter to warn about the dark future that awaits if we let ourselves be fooled is just one example of the way in which so many authors are seeking, in their interactions with LLMs, the evidence that will confirm that the right viewpoint has been achieved. 
Our built-in tendency to seek the Big Picture or the Whole Story when faced with an aesthetic object leads us to naturally want to ask questions like, What does it mean? How did we get here? Where is it going?, regardless of what specific prompt and output we are working with.  
Similar to how philosophers of art describe the correct (i.e., museological) staging of an artwork as
``both inside and outside history''~\citep[p.33]{preziosi2003brain}, there is a sense of being both outside of and part of the momentum of the technological progress: the spectator is in a position to recognize was has been learned from the past while simultaneously reflecting on its potential once deployed, which is typically assumed to be in the future. 
When we interact with a deep generative model like ChatGPT or DALL-E 2 it is as if time has frozen between the indeterminate past of the training data and the presumed future point at which the model is deployed toward some more specific goal in practice, and it is up to us at the present moment to find the meaning and resolve the end state toward which the technology is pointed.

But this ``urgency of perspective'' is not as unique to our current predicament with generative AI as we might like to think.  
If we accept that generative AI models staged for human interaction, like art, are treated as though they arise from a pre-determined principle or directive sense of purpose, then we should expect the moments in which the user stands before the output of a chatbot model to have a heightened status, just like the moments where the spectator gazes upon the work of art.
Some of the greatest works of art involve the spectator---literally through the gaze of the subject of the work, as in the Mona Lisa, or more metaphorically by defying the spectator's expectations, as in Duchamp's Fountain. 
At a higher level, the narrative of many art museums, like that of most natural history museums, presents modern man as the culmination of an evolving history. 
In the games of aesthetic judgment that we have so internalized, the feeling that there is a spotlight on the current moment is predictable. 
It feeds a pre-occupation with pricing AI today, despite evidence that our standards of judgment have evolved with the evolution of art.

Going a step further, if the evolution of the aesthetic object is taken as an expression of the continuously unfolding creative principle, 
then we would expect an aesthetic orientation to generative AI to direct our attention to the future.  
In the absence of the transcendental truth that we believe real artistic representation to reflect, the modal conclusion that generative AIs leave many of us with is that we are rushing toward dystopia.
So strong is the sense that AI is culminating toward something that the indefinite point in the future toward which AI marches has a name. The ``singularity'' is generally understood to be the point where artificial agents surpass human intelligence and cannot be controlled. Since it was first brought up by \citet{good1966speculations}, this concept has proven sticky \citep{chalmers2016singularity,kurzweil2014singularity}. It has enjoyed a kind of resurgence in recent attempts to philosophize generative AI, such as movements organized around aligning the values of artificial agents with human values in preparation for their increased agency in the future~\citep{kasirzadeh2023conversation,openai_alignment}.
As we shall see, the forward referencing implicit in our critiques of objects can be a strategy for positioning ourselves as somehow critical to the future trajectory of aesthetic production, even if we are merely the spectator rather than the artist.

\subsection{Universality, abstraction, and the erasure of differences}
In seeking the Big Picture or Whole Story, we are adept at overlooking how our seemingly neutral tendencies toward aesthetic judgment of objects also reflect social and political agendas. 
As an institution, the art museum has played a key role in establishing the centrality of a European perspective of the Other, the non-European cultures whose works of art were brought into the gaze of the museum and, by extension, the citizen of the European nation-state.
Through the creation of a mode of interpreting aesthetic objects based in abstraction and implied universality, the modern museum enacts an erasure of differences between different cultures, races, genders, and so on; ``a covering up of difference by a uniform visibility which de-others Others and domesticates all difference''~\citep{preziosi2003brain}. 
The introduction of the modern symbolic order epitomized by the museum is sometimes traced to the Crystal Palace, the large iron and glass structure created to house the Great Exhibition of the Works of Industry of All Nations in London in 1851. 
With its glass walls, transparent iron structure, and reliance on natural lighting throughout the structure, the Crystal Palace embodied the principles of this modern order: ``infinitely expandable, scaleless, anonymous; transparently and stylelessly abstract,'' 
offering a stage for ``making legible both the differences and similarities, and cognitive, aesthetic, and (consequently) ethical hierarchies among peoples by means of their juxtaposed and plainly seen products and effects'' \citep[pp.96--97]{preziosi2003brain}.

\begin{figure}
    \centering
    \includegraphics[width=0.5\linewidth]{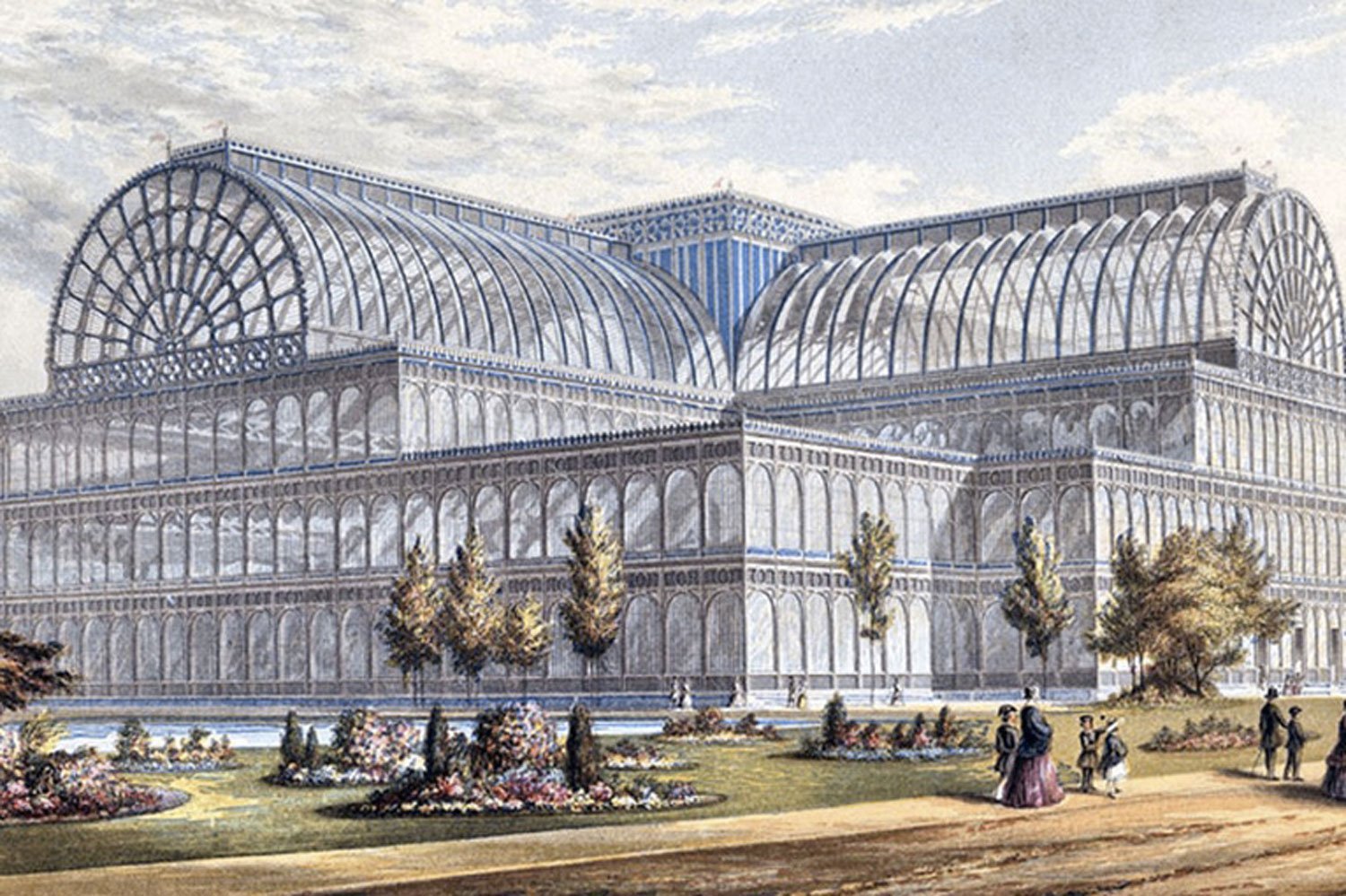}
    \caption{The Crystal Palace erected for the 1851 World's Fair, which introduced the ``opaque transparency'' associated with the symbolic order of the modern museum.}
    \label{fig:smiling}
\end{figure}

Similar critiques to those of the Crystal Palace as an ``opaque transparency'' founded on principles of abstraction and scale can be found in recent critical AI scholarship. 
The computer science communities that give us modern deep learning are criticized for prioritizing efficiency, scale, and generalization~\citep{birhane2022values}. 
The huge datasets on which progress has depended are accused of valuing ``efficiency at the expense of care; universality at the expense of contextuality; impartiality at the expense of positionality''~\citep{scheuerman2021datasets}. 
The everything-in-the-whole-wide-world benchmark---the idea that a dataset and metric can stand in for a much larger long-term goal like language or visual understanding---is deemed as fanciful as the idea of the muppet Grover in an 
``everything in the whole wide world'' museum~\citep{raji2021ai}.
Abstraction and independence from any particular content add to the allure of modern machine learning as a self-directed technology of vision.
Anecdotes about outputs of generative models that seem to epitomize de-othering are presented as evidence of the consequences of these values. ChatGPT distorts biographies to sound more typical; for example, replacing a community college with a more prestigious university.
DALL-E 2 generates images of Native Americans, Samurai soldiers, and African tribal warriors wearing wide grins, demonstrating how AIs dominated by American-influenced sources are producing ``a new visual monoculture of facial expressions'' that overwrites the varying cultural connotations of a smile~\citep{jenka2023}.

Ethicists and critiques advocate instead for a slower, more mindful approach to dataset curation, and a greater focus on specific tasks, locations, or audiences that embraces social and political influences on understanding the world~\citep{scheuerman2021datasets}.
One is reminded of movements like Arts and Crafts in art history, with its emphasis on traditional, anti-industrial values and methods. While the Crystal Palace captivated many across the world when it was unveiled, well-known proponents of the movement protested the exhibition on the grounds of its Machine Age values for creative production on a societal scale~\citep{willette2018}.

\begin{figure}
    \centering
    \includegraphics[width=0.5\linewidth]{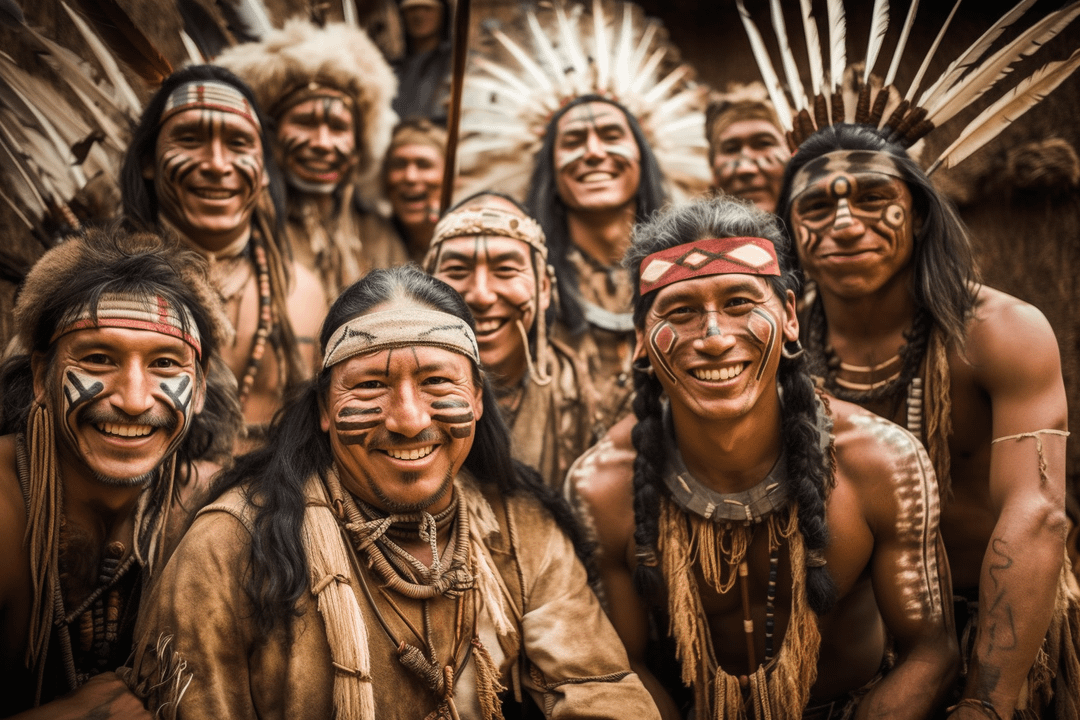}
    \caption{An image of smiling Native Americans generated by an AI~\citep{jenka2023}.}
    \label{fig:smiling}
\end{figure}

Ultimately, the Crystal Palace, like the modern art museum, points toward the future. 
If the goal is to secure a certain dominance of viewpoint, the centralization of perspective accomplished by modern museology can only be valuable if forward-pointing, so as to coordinate the intentions of the modern subject with the socio-cultural agenda engendered by the point of view~\citep{preziosi2003brain}.
There is, then, a built-in distancing to the system of organization implied by art history, which operates in a complementary way to the emphasis on the present mode.

Consider how leaders and policy makers are grappling with the question of how much attention should be given to preventing low probability but potentially catastrophic future events brought about by generative AIs versus to current known problems with the models~\citep{nature2023stop}.
Recent statements made by prominent researchers in AI about the risks of underdefined but dystopic future scenarios~\citep{pauseAI,statementAIrisk} have been critiqued for attempting to distance the work from modern day struggles with unfairness and other consequences of machine learning~\citep{gebru2023statement}. 
If we view this from a standpoint of how art historical practice has normalized the construction of an abstracted, omniscient, future-directed point of view focused on the fate of a generic human identity, this type of indefinite reference to the more distant future of society is not so surprising. 
It is a consequence of aestheticizing generative AI and we should expect it, like its art historical equivalents, to serve some social-political purpose.

\subsection{Echoes of divine terror}
So far, we have discussed various remnants of art historical modes of judging aesthetic objects that seem to persist in modern attempts to philosophize generative AI. 
Reading objects as signals of the mentalities of those who create them and as metaphors of the human essence, and expecting decorum and narrative and an abstracted human story from our interactions with generative AIs appear as symptoms of our ingrained tendencies toward aestheticizing objects. 
We have hinted at the purpose of these tendencies: they provide us with a reassuring illusion of legibility that centers our perspective and paints our current-day interactions as critical to an unfolding story of human aesthetic production.
But to really understand our temptations to engage in aesthetic judgment in the face of impressively human-seeming generative AI, we need to understand one more function aesthetic judgment has enacted. 
In the history of art, changes in aesthetic judgment have been most pronounced when we feel threatened by new modes of representation. 
To aestheticize human creative production is to restrict its power.

This takes some unpacking. To our modern sensibilities, a fear of art sounds ludicrous.
Plato's fear of the poet in the introductory quote is bewildering to the modern consumer of art who has become so accustomed to the idea of an art that has been tamed through art history, museology, and art criticism that it seems crazy to imagine seeing art as an active force in society the way we see technology, money, or politics. As \citet{nietzsche1996nietzsche} summarized, ``where do we see the influence, any influence, from art?''

But when we consider exactly what it would mean to be influenced by art (that is, influenced beyond mere aesthetic enjoyment or invoking of sentiment), or even for this to present itself as a real possibility, we observe similarities to our current historical moment. 
Plato feared that the poet's words would morally corrupt those who took them to heart, as if art could influence lives like alcohol or industrial technology or political propaganda or weapons of war. If this seems farfetched today, it is because in contrast to what we understand of classical Greece, our notion of human produced objects is split, separating the product from the work of art, and relegating the work of art to a sphere of aesthetic enjoyment.

\citet{agamben1999man} argues that in classical Greece 
there was no perceived difference between works of art and other man-made objects brought into being by ``technics'' or skilled activity. Manmade objects were distinguished only from natural objects, which ~\citet{aristotle_physics} described as containing in themselves their own principle and origin of entry into presence. 
With the industrial revolution and division of labor, however, this unitary status of objects not coming from nature was broken, leading to a distinction between
the work of art, which maintains permanent proximity to its origin and has the sense of completeness or being-at-work of a natural object, and the product, which does not possess its own principle of being, relying on people to instill the external model to which it conforms~\citep{agamben1999man}.
The work of art becomes associated with originality and authenticity, properties which preclude reproducibility. The formal principle of the product is simply the external paradigm that produced it, such that reproducibility is an essential property.

A work of art is then, compared to the product, more autonomous in a philosophical sense, yet essentially restricted from the principle of reproducibility that could render it useful in a worldly sense.  
To understand Plato's fear of moral corruption by the poet, we might imagine works of art ``that are in the world in the same way as the block of stone or the drop of water''~\citep{agamben1999man}, a ``Terrorist'' art where any sign is nothing but meaning~\citep{paulhan1941fleurs}.

We need look only as far as recent critical scholarship on generative AI to find allusions to outputs of models like ChatGPT that call to mind the terror of signs perceived as action or fact.
\citet{bender2021dangers} describe the dangers of LLMs 
whose outputs are taken at face value as human speech,. \citet{marcus2022jurassic} writes of how we should be terrified by AIs because they are unreliable or can hallucinate facts.

Seen in this light, reactions to generative AI today are intimately connected to the modern confusion and tendency for perversion when it comes to the distinction between the work of art and the product. 
The ready-made, like Duchamp's fountain, plays on the double-status of man's productive activity into art and product  by transferring an object from a technically reproducible state to one of aesthetic authenticity. 
Pop art, on the other hand, does the same but in reverse, from aesthetic status to industrial product \citep[p.63]{agamben1999man}.
These transitions are impossible given the mutual exclusivity of the essence of a work of art versus a product, but leave us with an aesthetically intriguing gesture toward a sort of availability-for-nothing status.

On the other hand, while the suspended museum-like quality of demonstrations of LLMs or generative vision models staged for consumption invoke our tendencies toward aesthetic judgment, they also threaten to overcome the inherent limitations on the ability of a work of art to have practical influence. 
Are the outputs of ChatGPT or DALL-E 2 reproducible? 
It would seem not. Yet they resist the authenticity or proximity to origin that allows us to confidently assert that some objects are art.
Generative models seem to be largely available-for-nothing now perhaps, but will this state continue? 
If and when they are put into the world to act as autonomous agents, how will our appraisals change? We are faced with an unfamiliar kind of indeterminacy in our aesthetic judgments, one that the ready-made or work of Pop art is prevented from inducing. 
An LLM may leave the sphere of the aesthetic at any time, and this is terrifying. 
It is no wonder that the moment of judgment---Does it have what it takes?---that is so familiar in art has also emerged in interactions with generative AIs, which seem to threaten our very livelihood.

\section{Why can't we put our finger on it?}

Once we recognize how generative AIs can incite our tendencies toward aesthetic judgment, we might wonder, where might these judgments take us?  
It is natural to seek the proper philosophical perspective through the third party commentaries of intellectual figures like Hofstadter, or the grandiose statements on AI risk by AI visionaries, or the cutting critiques of deep learning values by ethically-oriented scientists like Bender or Mitchell. 
But what, if any, traps might we encounter if we continue to operate with this orientation? 
We turn now to what questions we are NOT likely to answer if we let aesthetic judgment dominate our interactions with modern AI.   

\subsection{Ghost in the machine}
A tried and true means of stripping agency from the aesthetic object is to locate its truth in absence.
Philosophers and scholars of art history write about how aesthetic judgment, epitomized by taste in art, depends on negation~\citep{agamben1999man,kant1790, preziosi2003brain}. 
Similarly, critiques like Hofstadter's probe at the lack of a certain something that would betray humanness in ChatGPT's output, but without putting ever putting a  finger on exactly what that is. 
What is most exalted in art, and by extension in generative AI, is hard to define.
When we try to put our finger on what exactly it is that a person of taste can see in a work of art, just as when we try to summarize what it is that makes a generative AI intelligent, rarely do we arrive at something conclusive. 
As \citet{agamben1999man} summarizes, ``When we deny that a work is artistic, we mean that it has all the material elements of a work of art with the exception of something essential on which its life depends, just in the same way that we say that a corpse has all the elements of a living body, except the ungraspable something that makes of it a living being.''

Just as in our attempts to define the beautiful, it seems like our only recourse to identify intelligence is through its absence. 
The absolute standard used to be the Turing Test, but as economist and A.I. skeptic Gary Smith has pointed out, ``generating human-like conversation does not require or demonstrate intelligence in any meaningful sense of the word.'' By characterizing what computers and humans can and cannot do, the idea seems to be to move toward the production of intelligence that surpasses both. However, there's a potential no-true-Scotsman or ``God of the gaps'' situation, where intelligence ends up being defined as whatever computers cannot yet do.  As Hofstadter himself put it in his celebrated book, ``once some mental function is programmed, people soon cease to consider it as an essential ingredient of `real thinking' \dots AI is whatever hasn't been done yet''~\citep{hofstadter1999godel}.  We are like the art critic, who cannot remove himself from the piece, because it is only within his point of view, or gaze, that he finds his identity.

We assume a kind of smoothness to AI's behavioral space, that comes from our assumptions about people: that we can judge elements of what their internal representations must be by their actions, especially by their mistakes. We focus on examples for which our peers will agree that there is a right or wrong option, under the unproven hope that if an AI is correct where can verify it, it will be correct where can't verify it. But there is no reason to believe this about arbitrary function approximators, except to the extent we believe the way they smooth over the space of possible data has been demonstrated to have the same smoothness properties we expect from people. The brittleness of models with regard to their inputs suggests this isn't currently the case \citep{Zamfirescu-Pereira2023-ib}.

We might compare attempts at defining intelligence to Kant's definition of what is aesthetically pleasing as disinterested satisfaction, universality apart from concepts, purposiveness without purpose, and normality without a norm~\citep{kant1790}. 
As we have hinted above, what many don't realize is that such definitions empower us as spectators. So long as we can confine the determination of the worth of aesthetic objects to the realm of aesthetic judgment, we 
can continue to pursue 
breakthroughs of vision, works of representational ``magic'' without fear. 
Critics need not be able to explain how great works arose and cannot generally be expected to create great works themselves. 
However, through faculties of judgment, the critic puts the work of art in the realm of the aesthetic, so that he is essential to understand it. 
By introducing the critic as one who always points to the essence of the work by pointing beyond it, the system of art history establishes what theorists refer to as a deep ambivalence and sense of alienation in how we perceive art~\citep{preziosi2003brain}.

This tension is apparent in scholarly conversation as well. The researcher constructs batteries of tests to establish the intelligence of the generative model, each instantiating a particular hypothesis and constrained to a particular task, but its the lack of failures on these finite sets of tests that provides evidence. For each set of benchmark tasks that are said to establish, for example, ``sparks of artificial general intelligence'' in a program such as GPT-4~\citep{bubeck2023sparks}, critics are quick to point to how good performance on those tasks is ambiguous as a signal of intelligence. 
Some imply that the idea of creating a comprehensive benchmark capturing general intelligence is absurd, given ambiguities like the difference between the model generating something we understand versus understanding itself what it generated~\citep{mitchell_tweet}. 
Others seek to prove that creating systems with human-like or human-level intelligence is computationally intractable~\citep{vanrooij2023}.
What remains to be seen is if we can become convinced by such arguments, at least to an extent that we can curb our tendencies to fall into parochial seeming critiques.

\subsection{The genius versus the algorithm}
Another type of contradiction that arises when we lean on aesthetic judgment faculties instilled by our interactions with art works involves how we understand the human versus the algorithmic. 
Hofstadter's account of the story behind his book credits a conversation with his father, the contributions of books that inspired him, and the type-setting programs developed by a friend, but never for a minute do we doubt that the work reflects \textit{Hofstadter's} efforts, with the other characters playing bit roles. 
Another consequence of the development of the concept of taste in the history of art that we take for granted is that the figure of the artist becomes synonymous with a lone genius who is consumed by art. 
A source of irony in the perception of the artist that emerges is how with the rise of taste, it de-emphasizes any collaborative aspects of art. 
After the critic takes shape as a cultural figure, the artist is no longer seen as part of the community but instead is the ``individualistic creator pitted against the conforming masses''~\citep{montuori1995deconstructing}. Despite how many celebrated artists have relied on apprentices, from Rembrandt to Rodin, it's as if we're determined to ignore how for many artists and writers, finding ways to algorithmize and automate the creative process is an inherent part of the process.

Relatedly, demands for explanation and reproducibility seem opposed to the magic of the creative act, including that which we witness in interacting with a chatbot. 
Once something is fully explainable and reproducible, we cease to find it interesting, perhaps because it moves into the realm of product, hence relying on man for its external mode. The work of art is ``that which, even after one has achieved perfect knowledge of it, one is nonetheless unable to produce'' (\cite{kant1790}, as cited in \citet[p.24]{agamben1999man}).
If we imagine a generative AI that with outputs that are perfectly reproducible, the aura of interest is lost, much as how the creative act that leads to a great work of art must remain a bit mysterious to capture our attention.

One way to view demands for explanation is as an attempt to draw a line between humans and machines, in that while humans can be held accountable for acts, they are never required to explain where they come from. Even in a court of law, expressions such as ``I just did it, I don't know,'' occupy a million court transcripts.
If an AI can explain why it generated some output, can it be held accountable? 
Consider the creation of intentionally shocking contemporary artworks that comment on social or political realities, like scarecrows crouched to resemble migrant children separated from their parents at the border, or the remnants of a bombed police car from a riot presented as aesthetic object. 
Under what circumstances would we be comfortable calling a similar concept generated by an LLM and implemented as an installation in a museum a work of fine art? 
What would be the standard for the AI's rationale if it happened to ideate what might seem like an appropriate social commentary piece if it had been produced by a human?
While human creators are permitted to make vague artist's statements about the intentions of their work, leaving the real effort of explaining the piece's significance to the curator and critic, 
would we require a different definition of ``human enough'' if the piece originated with an LLM?

More broadly, we might say that human aesthetic is opposed to algorithmic reasoning, though considering how humans attempt to use algorithms or automation in their intellectual lives for a moment makes clear the problems with this view. 

Start with something simple such as washing dishes.  It's more pleasant to do this using a dish washing machine (setting aside whatever Zen-like pleasures one might get from a daily ritual of scrubbing).  But even in the absence of a machine, it is natural to want to algorithmize the process, to make it automatic.  Some of this is driven by the goals of saving time, physical effort, and natural resources, but it's not just that.  There's also a desire to reduce \textit{intellectual} effort. If we can effectively automate our dishwashing steps, then we can wash the dishes without thinking about it and instead devote our attention to more interesting or pleasant tasks such as conversation or listening to the radio.
In the dishwashing setting, we prefer to be more machine-like, and to the extent we engage our intelligence we prefer to do so by designing an efficient dishwashing plan, not by thinking hard about getting the dishes clean.

Next consider chess, that traditional arena of computer performance at a traditionally human task.
We play games for fun, but during a game most of us play to win. If we could win using an algorithm, many of us would do so.  To the extent that we are aware of such simple rules that work (``In the opening, move every piece once before moving any piece twice,'' etc.), we are inclined to follow them.  Of course, there's someone on the other side of that chessboard and there can be no automatic way to win, any more than there's an automatic way for opponent to win---but we are still trying to systematize what we do, even if in no other way than coming up with automatic ways of avoiding blunders.
That all said, if somehow it were possible to come up with an algorithmic way to win, or to guarantee a draw, then many of us would probably quit playing. The dishes need to get washed, but we have no reason to play chess if it's not interesting.

Finally, consider writing, which is more like dishwashing than chess in that it needs to be done, but different from either of those in that we want to use it to convey ideas and tell stories to the world that otherwise would not be told, or at least not shared so widely. Mark Twain is said to have always been trying to come up with a system to automate his writing.  
This may seem surprising because we think of Twain as a prolific creator, but he openly struggled with producing what he wanted, when he wanted it.  He wrote a lot, and wrote well, but never enough to satisfy his ambitions. 
The actual writing process is strenuous, in the same way that running laps is exhausting, and so it is a natural \textit{human} response to want to do it without such effort, whether that means recapitulating passages we've used in the past, copying structures we see other writers use well, or otherwise optimizing our process. 
The relationship between being human and using an algorithmic process is nuanced, whether we are talking about creative production or everyday chores. 
And so, we could say that many computer-generated works \textit{do} contain that ``je ne sais quoi'' that is familiar to us from human-created artworks, which is no surprise, given that (a) the computer generates the work by estimating parameters from, and thus abstracting from, corpuses of human-created art, and (b) when we, humans, create art, we often do it by a shuffling process, altering recombining existing ideas in a way not so much different from how these computer programs work.
Only on a plane imagined by our aesthetic judgments is what is human so separable from what is automatic.

\subsection{A conundrum of causal inference}
There is a final contradiction that arises when our interpretations of objects tolerate simultaneous 
metonymic and metaphoric readings, which we should expect if we prioritize aesthetic judgments of AI.
While stories of how AIs express values and emergent human qualities may be productive for driving scholarly interpretation and popular debate, 
they present us with a conundrum of causal inference.
Essays such as Hofstadter's represent attempts to debug, as it were, the false pretense of authenticity others might fall for when faced with the work of a generative AI.
But drawing the boundaries of what belongs in that story, and what does not, is a hopeless task. 
If the enabling assumption of art history and museology is that ``changes in form (or a lack thereof) are taken to correspond and reflect or embody changes (or a lack of change) in beliefs, attitudes, mentalities, or intentions, or to changes (or not) in social, political, or cultural conditions''~\citep[p.104]{preziosi2003brain}, 
then the aesthetic object can be read as a symptom of 
anything and everything that plausibly be thought to contribute to its appearance. 
At the same time, the object invokes a natural reading as emblematic of some universal essence. 
As a consequence, 
we find ourselves reliant on what appears to be a science of cause and effect, yet one that resists any attempts to fix a causal interpretation. 
For whenever we try to make a concrete argument about the cause behind some phenomena we see, such as the values of the cultural context in which it was created, 
we must acknowledge that it could also arise from that generic human essence or spirit or Mind~\citep[p.104]{preziosi2003brain}.
What aspects of Hofstadter's own account of his writing, we might ask, are reflections of the universal authenticity of human speech, versus words of his own choosing, informed by real world experiences he has had? 
We should expect no consensus on the answer, just like we should expect no consensus on the significance of the output of a model like ChatGPT.

Consider how even in the most carefully designed experiment to isolate what aspects of a modeling pipeline result in some behavior,
the possibility that it might be an expression of ``general intelligence'' cannot be ruled out without the experimenter bounding, through some set of assumptions, what constitutes general intelligence and in the process, seeming to contradict the potential for emergence. 
When our judgments of generative AIs fall back on archaic modes of interpretation developed for works of art, there is no end to the interpretations we can generate of what we see. But to what end do we produce these readings, when our ability to draw causal inferences is limited? 
This type of Catch-22 is likely to become more common with the rush to study and philosophize intelligence that we are now witnessing.

\section{What's next?}

\begin{displayquote}
``We believed \dots that man was a rational animal, endowed by nature with rights, and with an innate sense of justice; and that he could be restrained from wrong and protected in right, by moderate powers, confided to persons of his own choice, and held to their duties by dependence on his own will.'' --- \cite{jefferson1823}.
\end{displayquote}

Rationality has traditionally considered to be what separates us from the beasts, either individually (as in the above quotation) or through collective action, as in Locke and Hobbes.  More recently, though, it seems almost the opposite, that people are viewed as irrational computers. If the comparison point is a computer, then what makes us special is not our rationality but our emotions.

As we grapple with what it means to be human in light of machines that appear to mimic us and at times surpass our expectations of intelligence, 
we face a number of opportunities (or, seen in a different light, traps) in how we orient ourselves to the outputs of generative AIs.
Above we have argued above that the ways we seek meaning from our interactions with generative AI enact the same form of beauty contests we have long relied on in the face of aesthetic objects. 
In our attempts to distance and empower ourselves in the face of objects that blur the distinction between art and product, we become reliant on a negative theology to define success. 
We accept as neutral modes of reading and taxonomizing objects that have always had social and political undertones. 
We remain devoted to an idiosyncratic definition of creative genius that precludes acknowledging the collaborative and algorithmic aspects of great works of human creative production. 
Ultimately, we find ourselves unable to find solid enough footing to generate any causal inferences with confidence, despite our desire for perspective.

One implication of our analysis is that our appraisals of the significance of the latest progress in generative AI are unlikely to ``price in'' even in the near future with the same value that we believe them to hold today. 
We should expect the nature of our aesthetic judgments to change with the technology; in the history of art, this has always been the case. As long as generative AI outputs remind us of the creative expressions of humans, the nature of our judgments is likely to evolve.  But how exactly our judgments will change with the technology is an open question.

In many ways, the current moment can be seen as the next step in a historical trajectory where the power of the spectator has grown and the artist faces an increasingly futile task in trying to find some identity in his work. 
The rush to philosophize generative AI may be a response to changes in the scale of aesthetic production, similar to what occurred upon the explosion of creative work in the Renaissance era due to innovations in art and publishing and greater spread of ideas with trade. 
The notion of taste in art arose later in this period, perhaps as a sort of necessary precaution, sort of like the three-quarter closing of a photo lens in the wake of a  bright object~\citep{agamben1999man}. This was followed by the creation of the curated perspective of the modern museum.
As the concept of taste has evolved, the power of the critic in determining the final significance of art has only grown, and a return to the state that preceded the rise of the critic, where the status of art was not so highly secularized, has been deemed impossible by philosophers~\citep{agamben1999man,hegel1835lectures}.

Generative AIs threaten to again blow up the scale of aesthetic production by``reducing the cost of generating disinformation to zero''~\citep{marcus2022jurassic}, fueling our sense of urgency at putting them in their place. 
From a certain perspective, generative AI removes the need for an artist to generate the art. If they want to, spectators and critics can do it themselves.

What happens when the critic embraces this newfound ability? This is the scenario that gives us AI-generated music, for instance, where savvy producers have pawned off synthetically-generated songs using the voices of Drake and Frank Ocean as the real thing~\citep{hoover2023music}.   
What is not clear, however, is what limits there may be on optimizing AI art to give the spectator exactly what they want. 
For example, can we hope to statically achieve a state in which generative AIs give many people exactly what they want, eliminating our urge to point to a gap? It seems more likely that our judgment of creative production are too fluid and historically situated to reach this point.

But how long can we expect the current situation---in which most AI does not produce the sort of aha moments by which one realizes something about their own experience or thoughts or perspective due to something the AI did---to last?
We can liken our position at present to judging a child prodigy who is unexpectedly technically capable but lacks recourse to the experience of life to make something that strikes us. It remains difficult, however, to rule out the possibility that the aha moment may be just a few orders of magnitude of compute and data away.
And so, a sense of the fear of this moment, when ``unquestionably insightful people'' will fall under the influence, is palpable in recent criticism of generative AI and seems unlikely to go away anytime soon.

Part of this fear may be that generative AI will be able to hack the human psyche, like a visual illusion hacks the visual perception system, and that this possibility might emerge as a form of deception to which we are particularly vulnerable. 
We can compare this fear to the art critic's fear of bad taste, which also has a storied history. 
During the 17th century, the type of man who had bad taste comes into intellectual focus as the person who 
loves ``what is short of the right point or beyond it,'' does not know how to identify the perfection of the work of art by distinguishing truth from falsehood~\citep{agamben1999man}.
The more the cultural figure of the man of taste seemed to purify his aesthetic faculty, the more he seemed to desire its opposite, as if good taste wants to pervert itself every chance it gets. 
Might we be afraid of our own interest in what a generative AI, trained to respond to human feedback, might create for us?

There is an art---or perhaps a science---to manufacturing experiences that are generic and formulaic yet with just enough disturbance or surprise to keep us interested.    
Consumers of art often seem to want such experiences, such as from TV shows with generic heist, romance, family, or cop show themes, which appear to aim for the asymptote of machine production. Much has been written over the years on the balance between predictability and surprise in music and literature \citep{huron2006,byrne2012,tobin2018}. An important role of much art is not to tell a particular person's story or achieve some aesthetic height but rather to provide a pleasant confection with plot twists that are expected in their structure but not their details, instant-classic riffs that sound just right while being entirely new.  
Fear of a future in which content tailored to achieve just the right mix of camp or sex or heartbreak or visual extravaganza for crowds who seem indifferent to good taste may well explain some of the horror around generative AIs. 
That recent fear of generative AI is just the latest incarnation of fear of bad taste suggests that much of society's tacit agreement on taste has been underwritten by the inability to test its boundary conditions.
Our point here is not to disparage mass taste but rather to recognize that human effort and creativity, however defined, can often be seen as means rather than ends in the creation of art.
The blurry line between art and product helps fuel the perception that as a society, we need taste.

It is easy to imagine generative AIs becoming a kind of art historical tool, one that puts aesthetic judgment itself, with all of its pretenses and perversions, under the spotlight. 
Off-putting synthetic photos of people of varying cultures all grinning in exactly the same way become objects of study, and this same approach can be applied beyond visual media, to characterize human culture as expressed in literature, poetry, music, even sculpture~\citep{diaz2023}.
In addition to driving the cost of generating misinformation to zero, generative AI might offer a more direct, synthetic and at-scale version of the work of artists like Walton Ford, Honoré Daumier, John Heartfield, or good cartoonists who have used visuals to caricature human culture and expose its deepest faults.

But who will be blamed for the content that gets produced in bad taste, the ``artist'' or the critic? 
Authorship was cemented when it was enshrined in law via concepts such as copyright, to allow for punishment for transgressions \citep{foucault2017author}. But if a work seems to come from nowhere, because generative AI is not granted standing as a subject, who is to be punished when responsibility becomes too diffuse to make sense of with current metaphors?

Also related to ``wireheading'' are questions of how much the experience of a given generative model will become individualized. At present it remains expensive and time consuming enough to create one's own model that most do not. However, this is rapidly changing. 
Consider an infinite sitcom that (with overwhelming likelihood) wouldn't repeat for thousands, if not millions of years of viewing. Take this to the extreme: imagine a TV show with a massive cast, all virtual, that puts out media at the rate of 100 years of content a second. Will people settle on sections of the media to congregate around? To what extent will the experience of media be (intentionally or unintentionally) individualized?   We can imagine a self-contained loop in which AI-created entertainment is directly reviewed by AI-trained critics.

Currently the unified nature of ChatGPT's apparent consciousness results from a trick of the user interface and the expense of the creation of the system. But if this illusion crumbles, will the selection of certain elements into a complete package become the real act of value? 
Think of the earth as a piece of art, on top of which governments have placed parks to highlight certain aspects that are deemed worthy of being their own museum. Perhaps this will be the new way we think about AI: as a natural process we curate for our ends, as a resource which we organize because it is too cheap to meter.

Does the science of generative artificial intelligence need the equivalent of professional art critics, similar to the value of professional critics of science in general \cite{Card2020-tb}? Or is this a temporary moment because our interactions with state-of-the-art AI occur through ``chats'' that are mostly removed from the rest of life (similar to an art museum)? Whatever the case may be, we should expect our tendency to fall back on aesthetic judgment to carry philosophical consequences.

\section*{Acknowledgments}
The characterization of art history and the philosophy of aesthetics on which these arguments are based was introduced to the first author by Donald Preziosi during roughly a year of graduate study under his advising. 
We also thank Douglas Hofstadter for helpful feedback.

\bibliography{ref}

\end{document}